\documentclass[prl,twocolumn,english]{revtex4-1}

\usepackage{array}
\usepackage{amsmath,amssymb}
 \usepackage{tikz-feynman} 
\usepackage{graphicx}
\usepackage{subfigure}
\usepackage{mathrsfs}
\usepackage{bm}
\usepackage{mathtools}
\usepackage{epstopdf}
\begin{document}
\title{Bipolarons in a Bose-Einstein condensate}
\author{A.\ Camacho-Guardian}
\affiliation{Department of Physics and Astronomy, Aarhus University, Ny Munkegade, DK-8000 Aarhus C, Denmark}
\author{L. \ A. \ Pe{$\tilde{\rm n}$}a Ardila}
\affiliation{Department of Physics and Astronomy, Aarhus University, Ny Munkegade, DK-8000 Aarhus C, Denmark}
\author{T. \ Pohl}
\affiliation{Department of Physics and Astronomy, Aarhus University, Ny Munkegade, DK-8000 Aarhus C, Denmark}
\author{G. \ M. \ Bruun}
\affiliation{Department of Physics and Astronomy, Aarhus University, Ny Munkegade, DK-8000 Aarhus C, Denmark}
\begin{abstract}
Mobile impurities in a Bose-Einstein condensate  form quasiparticles called polarons. Here, we show that two such polarons  can bind to form a bound bipolaron state. Its emergence is caused by an induced nonlocal interaction mediated by density oscillations in the condensate, 
and  we derive using field theory an effective Schr\"odinger equation 
describing this for arbitrarily strong impurity-boson interaction. We furthermore compare with Quantum Monte Carlo simulations  finding remarkable agreement, which underlines the predictive power of the developed theory. It is found that bipolaron formation typically requires strong impurity interactions beyond the validity of more commonly used weak-coupling approaches that lead to local Yukawa-type interactions.
We predict that the bipolarons are observable in present experiments and describe a procedure to probe their properties.
\end{abstract}
\date{\today}

\maketitle

The notion of quasiparticles is a powerful concept that is indispensable for our understanding of a wide range of problems from Helium mixtures and condensed matter systems to nuclear matter~\cite{BaymPethick1991book,Mahan2000book,ring2004nuclear}. 
Quasiparticles can experience induced  interactions mediated by their surrounding. The induced interaction is inherently attractive and can therefore 
lead to the formation of bound states. This is the origin of Cooper pairing in conventional superconductors~\cite{Cooper1956} where 
 the size of the Cooper pairs typically is 
 much larger than the average distance between unbound quasiparticles. Bipolarons stand out as an important example of the opposite limit, where two quasiparticles, so-called polarons, form a bound state much smaller than the average distance between the 
unbound polarons.  The formation of bipolarons is suggested to be the mechanism behind electrical conduction in polymer chains~\cite{Bredas1985,Mahani2017}, organic magnetoresistance~\cite{Bobbert2007}, and even high temperature  superconductivity~\cite{Alexandrov1994,Alexandrov2008}. 

The recent experimental realisation of polarons in ultracold quantum gases~\cite{Schirotzek2009,Kohstall2012,Koschorreck2012,Cetina2016,Scazza2017,Jorgensen2016,Hu2016} has opened up unique opportunities  to study the quasiparticle physics  in a  highly controlled manner. So far, experimental and theoretical efforts have focused 
on single-polaron properties in degenerate Fermi~\cite{Schirotzek2009,Kohstall2012,Koschorreck2012,Cetina2016,Scazza2017} and Bose gases~\cite{Jorgensen2016,Hu2016}, for 
which we now have a good understanding.  
Bipolarons in Bose Einstein condensates (BECs) have been explored within the Fr\"ohlich model \cite{Mahan2000book}, which is valid only for weak interactions~\cite{Casteels2013}. 
Yet, their observability hinges on sufficiently strong binding, and the formation of  bipolarons in atomic gases   remains an outstanding question that requires a new 
theoretical framework for strong interactions.

\begin{figure}[!h]
\begin{center}
\includegraphics[width=\columnwidth]{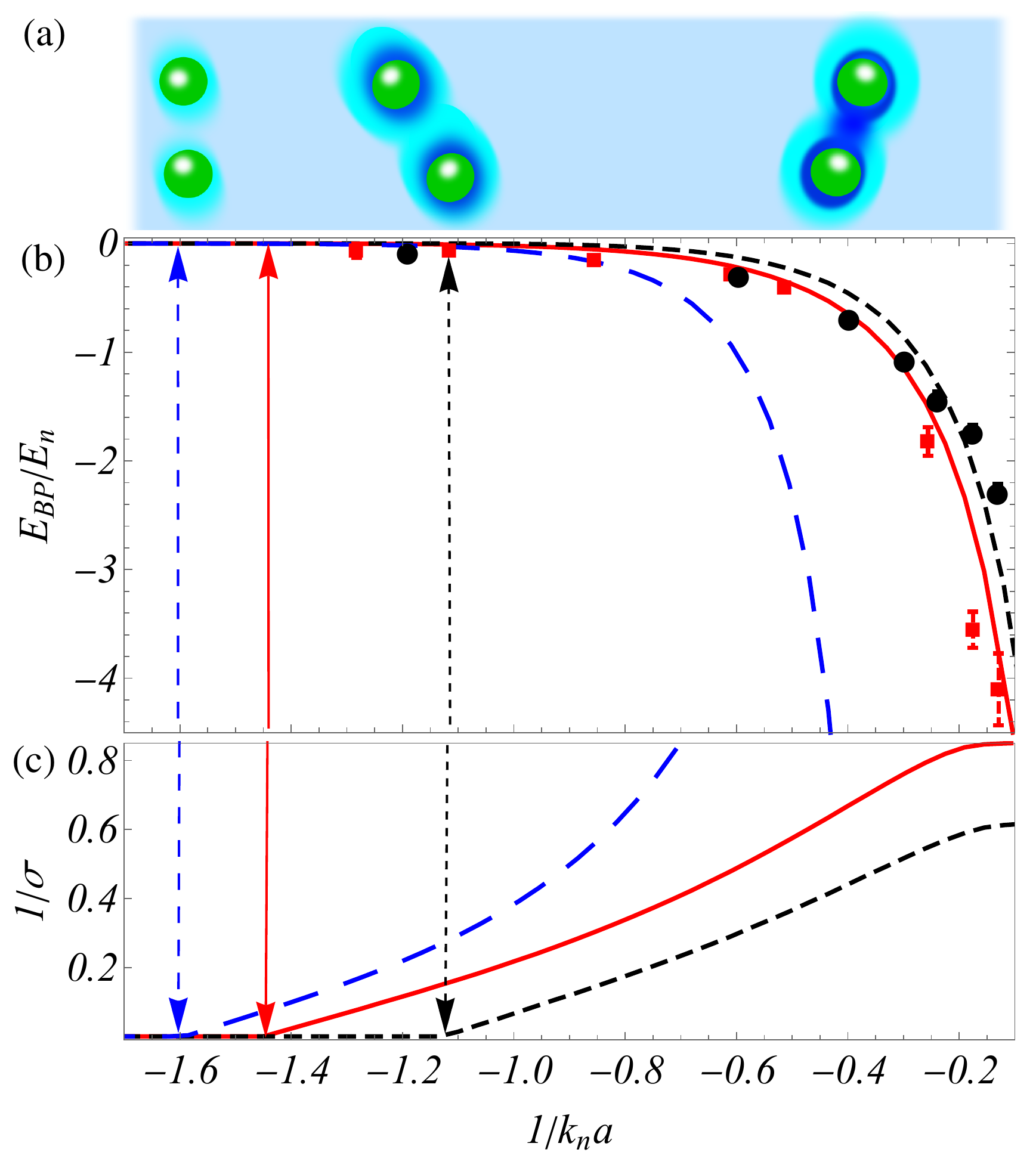}
\end{center}
\caption{(a) The cartoon shows Bose polarons forming a bipolaron as a consequence of a mediated interaction. (b) Binding energy $E_{BP}$ of  the bipolaron  as a function of the impurity-boson interaction strength for two bosonic impurities with $m=m_B$. The red solid and 
black dashed lines are solutions to Eq.\ \eqref{Schrodinger} with the induced interaction given by Eq.\ \eqref{Veff} for the  gas parameters $n_Ba_B^{3}=10^{-6}$
and $n_Ba_B^{3}=10^{-5}$. The red squares and black circles are the results of the DMC calculations for the same two gas parameters.   The 
blue long dashed line is the ground state energy of the  Yukawa interaction Eq.\ \eqref{Veffweak} for  $n_Ba_B^{3}=10^{-6}$. 
(c) The corresponding inverse size $1/\sigma=\xi_B/\sqrt{\langle r^2\rangle}$ of the bipolaron wave function, where $\xi_B=1/\sqrt{8\pi n_Ba_B}$ is the BEC coherence length. Vertical arrows denote the critical strength to form a bound state.  
}
\label{EnergyBoseBipolaron}
\end{figure}

In this Letter, we present such a theory and demonstrate that two impurities immersed in a BEC can indeed form bound states for sufficiently strong interactions between the impurities and the condensate atoms. Based on field theory, we derive an effective Schr\"odinger equation with a \emph{nonlocal} polaron-polaron interaction that describes the emergence of bipolarons. This effective description provides an intuitive and feasible approach to account for arbitrarily strong impurity-boson interactions, and it is 
furthermore shown to be in remarkable agreement with first-principle   quantum Monte-Carlo results. Our  theory allows to reliably predict the existence of bipolarons under  realistic conditions, and demonstrates that it is possible to realise bipolarons with sufficiently strong binding to enable their observation.

We consider two impurities of mass $m$ immersed in a zero-temperature BEC of bosons with mass $m_B$ and density $n_B$. As typical for cold-atom experiments, the BEC features weak interactions with  $n_B^{1/3}a_B\ll 1$, so that it is accurately described by Bogoliubov theory. Here, $a_B$ is the scattering length for the zero-range boson-boson interaction. The interaction of a single impurity with the BEC is characterised by the scattering length $a$, and it 
leads to the formation of the Bose polaron~\cite{Li2014,Levinsen2015,Shchadilova2016,Tempere2009,Rath2013,Christensen2015,Grusdt2017,Astrakharchik2004,Cucchietti2006,Ardila2015,Ardila2016},
which was recently observed experimentally~\cite{Jorgensen2016,Hu2016}.
 
\begin{figure}
\begin{center}
\includegraphics[width=0.9\columnwidth]{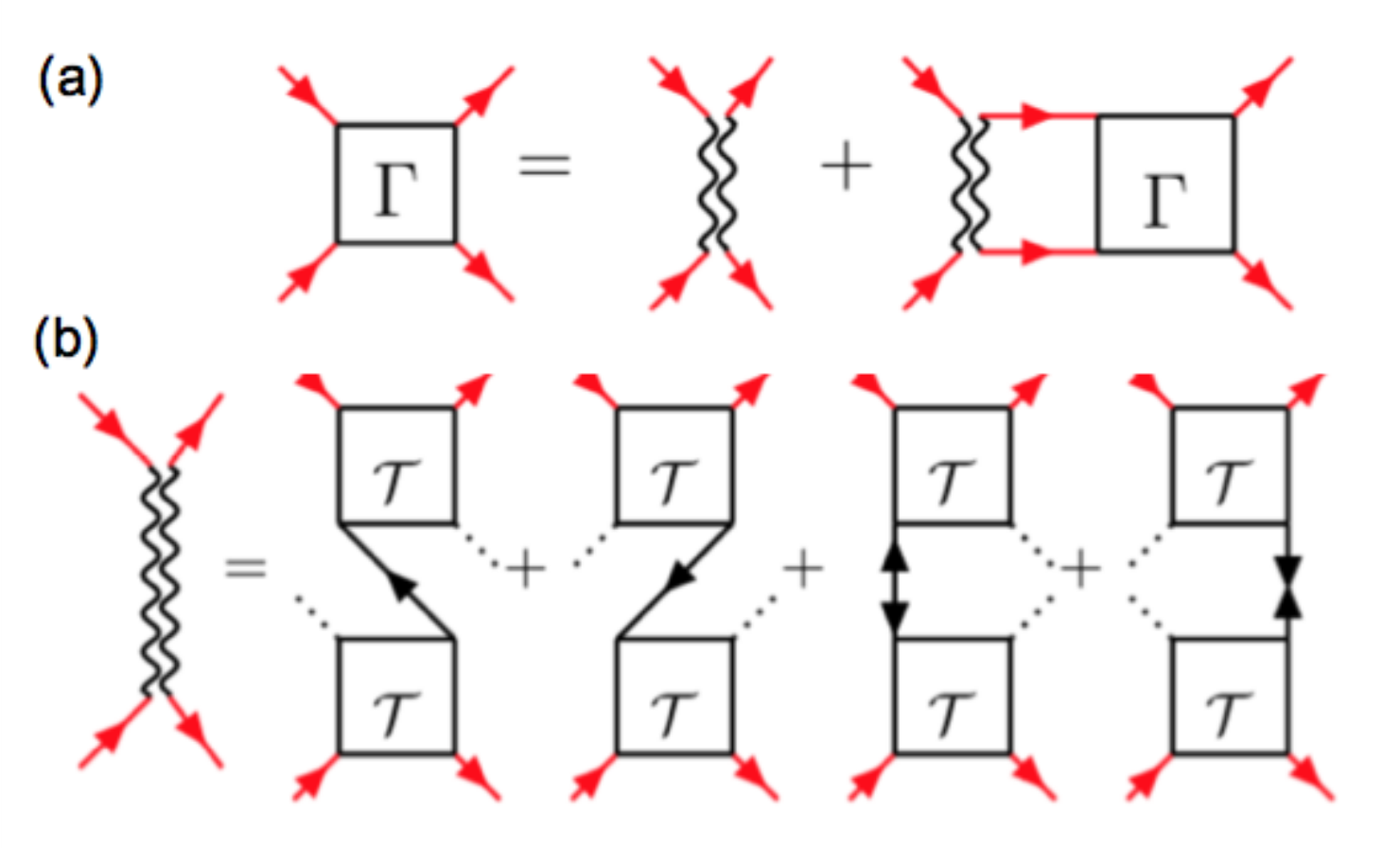}
\end{center}
\caption{(a) Diagrammatic representation of the Bethe-Salpeter equation for  impurity-impurity scattering.
Red lines are the impurity Green's function and the double wavy line is the induced interaction. 
(b) The induced interaction. Black lines are normal and anomalous BEC Green's functions, dashed lines are condensate bosons, and $\mathcal{T}$ 
is the impurity-boson scattering matrix in the ladder approximation.  }
\label{BetheV}
\end{figure}

Two polarons can interact strongly by exchanging density fluctuations in  the BEC, even when there is no significant direct interaction between the actual impurities. This induced interaction is inherently attractive and can therefore facilitate bound dimer states, as illustrated in Fig.~\ref{EnergyBoseBipolaron}(a).
 Within a field-theoretical formulation, two-body bound states in a quantum many-body system can be identified as poles of the generalised scattering matrix $\Gamma$. 
 Considering the scattering of two impurities from states with energy-momenta  $({ k}_1,{ k}_2)$  to   $({ k}_3,{ k}_4)$, the Bethe-Salpeter equation for the scattering matrix 
  reads in the ladder approximation~\cite{Fetter1971} [see Fig.\ref{BetheV}(a)] 
\begin{gather}
\Gamma({ k}_1,{ k}_2;{k}_1-{ k}_3)=V({ k}_1,{ k}_2;{ k}_1-{ k}_3)+\sum_{q}V({ k}_1,{ k}_2;q)\nonumber \\
\times G({ k}_1-q)G({ k}_2+q)\Gamma({ k}_1-q,{ k}_2+q;{ k}_1-q-{ k}_3).
\label{BetheS}
\end{gather}
Here $G({ k})$ is the impurity Green's function, $k=(\mathbf k,z)$ is the four momentum vector, and $V(k_1,k_2;q)$ is the induced  interaction between two impurities. 
We calculate this interaction using the diagrammatic scheme illustrated in Fig.~\ref{BetheV}(b),  which 
simultaneously accounts  for arbitrarily  strong boson-impurity scattering and the propagation of density waves in the BEC~\cite{Camacho2017,Suchet2017}. 

In order to derive an effective Schr\"odinger equation for the bipolaron, we change our description 
from bare impurities to polarons by approximating the impurity Green's functions in Eq.\ \eqref{BetheS} by their 
 value around the polaron poles, i.e.\
  $G(k)\simeq Z_{\mathbf k}/(z-\omega_{\mathbf k})$. Here $\omega_{\mathbf k}$ is the energy of a polaron 
 with momentum ${\mathbf k}$ and quasiparticle residue $Z_{\mathbf k}$. Here we consider $\hbar=1$.
 We furthermore multiply the Bethe-Salpeter equation \eqref{BetheS} by $Z_{\mathbf k_1}Z_{\mathbf k_2}$
 so that it gives the scattering matrix $\Gamma_P$ of two polarons instead of two impurities. This gives 
 \begin{equation}\label{eq:Veff}
V_{\text{eff}}( k_1, k_2;{ q})=Z_{\mathbf {k}_1}Z_{\mathbf {k}_2}V( k_1, k_2;{ q})
\end{equation}
 for the effective polaron-polaron interaction. Since it depends on the incoming $k_1$ and $k_2$, as well as the transferred four-momentum $q$, 
 a direct solution of the Bethe-Salpeter equation is very difficult. We therefore neglect retardation effects and take the static
  limit of the interaction  setting all energies to zero in $V_{\rm eff}$. 
  This is a good approximation if the binding energy $|E_\text{BP}|$ of the bipolaron is smaller than the typical energies of 
  the Bogoliubov modes exchanged between the polarons, i.e.\ 
  if $\sqrt{|E_\text{BP}|/m}\ll c$ with $c=2\sqrt{\pi n_Ba_B}/m_B$ the speed of sound in the BEC. Neglecting the frequency dependence of $V_{\text{eff}}$ means 
  that the frequency sum involving the two impurity Green's functions in Eq.\ \eqref{BetheS} can be performed analytically. 
The Bethe-Salpeter equation \eqref{BetheS} then reduces to the Lippmann-Schwinger equation, which in turn is equivalent to the
Schr\"odinger equation for two polarons  interacting via an instantaneous interaction. It reads in the center of mass (COM) frame
\begin{align}
E_\text{BP}\psi(\mathbf k)=2\omega_{\mathbf k}\psi(\mathbf k)+\sum_{\mathbf k'}V_{\rm eff}(\mathbf{k},\mathbf{k}')\psi(\mathbf k'),
\label{Schrodinger}
\end{align}
where  $\psi(\mathbf k)$ is the relative wave function of the bipolaron with energy $E_\text{BP}$. The effective interaction for 
two polarons with momenta $(\mathbf k,-\mathbf k)$ scattering into  $(\mathbf k',-\mathbf k')$ is 
\begin{gather}
V_{\rm eff}(\mathbf{k},\mathbf{k}')= Z^2n_B\left[2\mathcal{T}(\mathbf{k},0)\mathcal{T}(\mathbf{k'},0)G_{11}(\mathbf{k}-\mathbf{k}',0)\right.\nonumber\\
\left.+\mathcal{T}^2(\mathbf{k},0)G_{12}(\mathbf{k}-\mathbf{k}',0)+\mathcal{T}^2(\mathbf{k'},0)G_{12}(\mathbf{k}-\mathbf{k}',0)\right]
\label{Veff}
\end{gather}
where  $G_{11}({\mathbf k},0)$ and $G_{12}({\mathbf k},0)$ are the normal and anomalous Green's functions for the bosons, and 
$\mathcal{T}(\mathbf{k},0)$ is the boson-impurity scattering matrix, all evaluated at momentum $\mathbf k$ and zero energy. 
Note that $\mathcal{T}$ is distinct from $\Gamma_P$, which describes the scattering of two polarons.
We calculate  the polaron energy $\omega_{\mathbf k}$ and residue $Z_{\mathbf k}$  using an extended ladder scheme  with the 
 effective mass approximation $\omega_{\mathbf k}=\mathbf{k}^2/2m^*+\omega_0$, where $\omega_0$ is the energy of a zero momentum polaron, and 
  assuming that $Z_{\mathbf k}\approx Z_{\mathbf k=0}$. This scheme agrees well both with experimental data and with Monte-Carlo calculations 
  for the single polaron properties. More 
 details  are given in the Supplemental Material~\cite{SM}.

With Eq.\ \eqref{Schrodinger}, we have arrived at an effective  Schr\"odinger equation for the bipolaron. In addition to providing an intuitive picture, it is much simpler to solve 
than the full Bethe-Salpeter equation \eqref{BetheS}, yet it gives accurate results even for strong coupling as we shall demonstrate
 shortly. The fact that Eq.~\eqref{Schrodinger}
is a two-body effective description of an underlying many-body problem is reflected in the energy dispersion $\omega_{\mathbf k}$ and 
by the fact that the interaction is \emph{non-local}, i.e.\ 
 $V_{\rm eff}(\mathbf{k},\mathbf{k}')\neq V_{\rm eff}(\mathbf{k}-\mathbf{k}')$. It  becomes local only for weak coupling $|k_na|\ll 1$ with $k_n^3/6\pi^2=n_B$, 
 where the boson-impurity scattering matrix reduces to the constant $\mathcal{T}_\nu=2\pi a/m_{\text{BI}}$ with $m_{\text{BI}}=mm_B/(m+m_B)$. 
Equation \eqref{Veff} then simplifies  to the well-known second order (in $a$) Yukawa expression    
\begin{align}
V_{\rm eff}(\mathbf{k},\mathbf{k}')=-\mathcal{T}_\nu^2\chi(\mathbf k-\mathbf k',0),
\label{Veffweak}
\end{align}
  where $\chi(\mathbf k,z)=n_B k^2/[m_B(z^2-E_k^2)]$ describes density-density correlations in the BEC. 
Our theory extends this result into strong  coupling by including multiple impurity-boson scattering. 

\begin{figure}
\begin{center}
\includegraphics[width=\columnwidth]{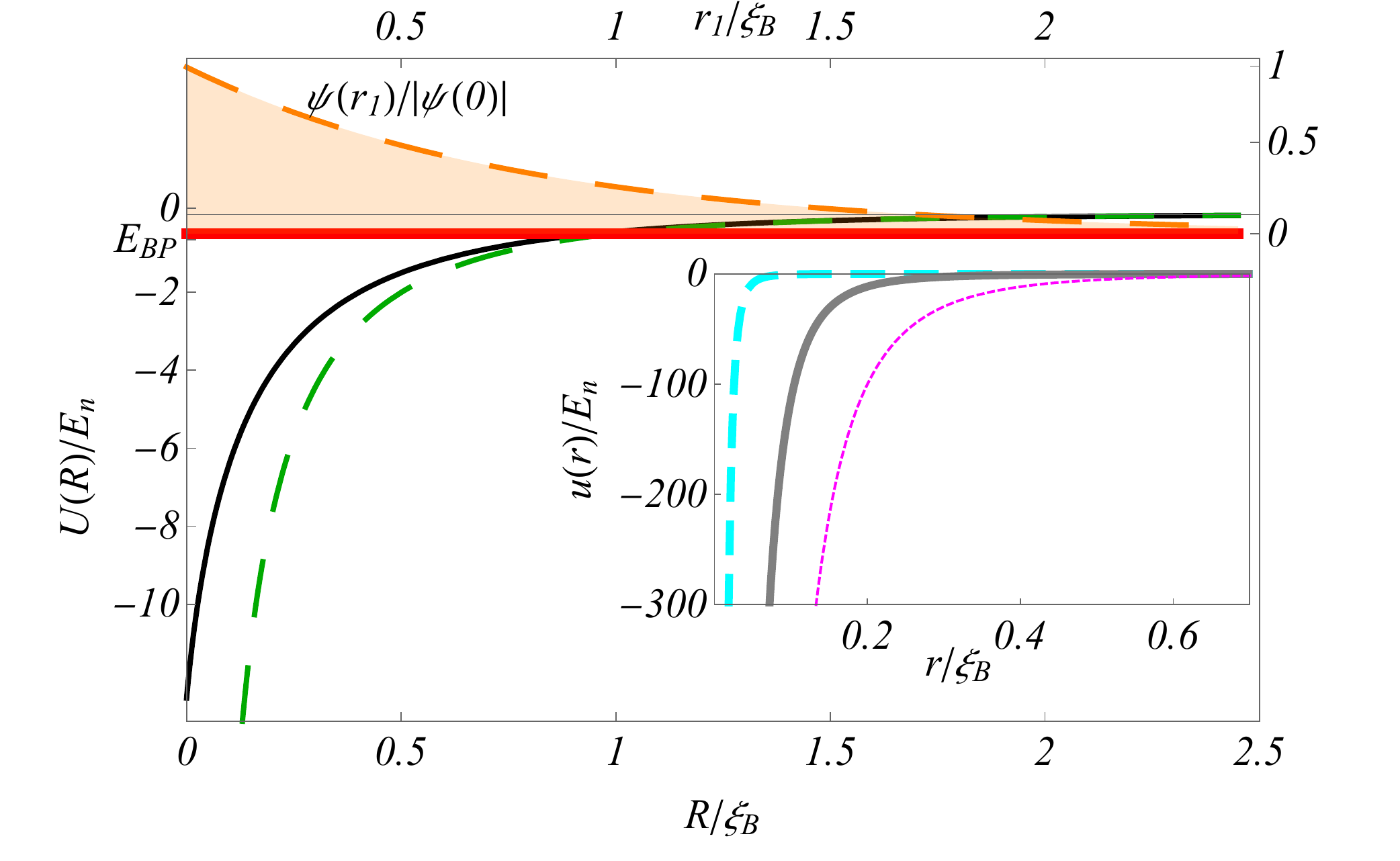}
\end{center}
\caption{The local part $U(R)$ of  $V_\text{eff}({\mathbf r}_1,{\mathbf r}_2)$ (black solid) and 
the Yukawa interaction (green dashed) for $n_Ba_B^3=10^{-6}$ and $1/k_na=-0.4$. The corresponding $s$-wave binding energy $E_\text{BP}$ and wave function 
are shown by red solid and dashed orange lines. Inset:  the non-local part $u(r)$ for $1/k_n a=-10$ (dashed blue), $-1.5$ (solid gray), and $-0.4$ (short dashed purple). }
\label{NonLocalPotential}
\end{figure}

We notice that in real space, the non-local interaction term in Eq.\ \eqref{Schrodinger} reads
 $\int\! d^3r_2V_{\rm eff}({\mathbf r}_1,{\mathbf r}_2)\psi({\mathbf r}_2)$.
To  quantify the non-locality, we  write  $V_{\text{eff}}(\mathbf r_1, \mathbf r_2)$ as a function of  $\mathbf r =\mathbf r_1-\mathbf r_2$ 
and $\mathbf R =(\mathbf r_1+\mathbf r_2)/2$, where $\mathbf r_1$ and $\mathbf r_2$ denote the relative distances between the in- and out-going polarons. 
The local Yukawa interaction Eq.\ \eqref{Veffweak} can then be written as 
$V_{\text{eff}}(\mathbf R, \mathbf r)=\delta(\mathbf r)\alpha\exp(-\sqrt2 R/\xi_B)/R$ in real space, where 
 $\alpha={\mathcal T}_\nu^2n_Bm_B/\pi$. 
 We define the "local" and "non-local" parts of the interaction  as  $U(\mathbf R)=\int\! d^3r V_{\text{eff}}(\mathbf R, \mathbf r)$ and  $u(\mathbf r)=\int\! d^3 R V_{\text{eff}}(\mathbf R, \mathbf r)$.  For the Yukawa interaction, we have $U(\mathbf R)=\alpha\exp(-\sqrt2 R/\xi_B)/R$ and 
$u(\mathbf r)\propto\delta(\mathbf r)$.  Figure \ref{NonLocalPotential}  plots  $U(\mathbf R)$ for  $n_Ba_B^3=10^{-6}$ and $1/k_na=-0.4$. 
We see that whereas $U(\mathbf R)$  approaches the Yukawa form for large distances, 
it it differs significantly  for $R/\xi_B\lesssim1$. In particular, $U(\mathbf R)$ is finite for $R\rightarrow 0$. We also plot the wave function $\psi({\mathbf r}_1)$
of the lowest bound state off-set vertically by its  binding energy $E_\text{BP}$, to illustrate that it 
extends well beyond the classical turning point $U(R)=E_\text{BP}$. This is a consequence of the non-local character of the interaction. The inset of 
Fig.\ \ref{NonLocalPotential} plots $u({\mathbf r})$, which shows that the non-locality given by the width of $u({\mathbf r})$
increases with increasing interaction. This non-locality is a characteristic sign of the underlying many-body physics, which is  analogous to the case of the  nuclear force~\cite{Machleidt1996}.

In order to verify the accuracy of our theory and the involved approximations, we also perform diffusion Monte-Carlo (DMC) simulations \cite{Ardila2015,SM},
 which in principle takes into account all possible impurity-boson correlations.
  To this end, we determine the ground state energy, $E_0$, for a BEC of $N$ particles in a box with periodic boundary conditions. We then obtain the bipolaron binding energy 
$E_{BP}=E-2\omega_0=E_2-2E_1+E_0$ from the ground state energies $E_1$ and $E_2$ of the same condensate but containing one impurity and two impurities,
 respectively. Details of the DMC calculations are given in the Supplemental Material~\cite{SM}.

Figure \ref{EnergyBoseBipolaron}(b) shows the bipolaron binding energy $E_{BP}$ in units of $E_n=k_n^2/2m$
as a function of the impurity-boson scattering length $a$. We consider the case of bosonic  impurities, so that the bipolaron wave function is symmetric under particle exchange ($s$-wave symmetry). Results obtained from  our DMC simulations and the effective Schr\"odinger equation using the  interaction
Eq.\ \eqref{Veff}  as well as Eq.\ \eqref{Veffweak} are compared for two different BEC gas parameters. We keep $a<0$ here and in the following. For both  interactions, we find that bound bipolaron states with $E_{BP}<0$ emerge beyond a critical interaction strength, $k_na_c$, which is marked by the vertical lines in Fig.\ \ref{EnergyBoseBipolaron}.
Beyond this critical value, the binding energy initially increases very slowly, since the  polaron-polaron interaction is at least a second order effect in $a$.  
For stronger coupling $k_n|a|\gtrsim1$, the binding energy crucially becomes significant
compared to the single-polaron energy $\omega_0$, which is maximally of order 
$E_n$ \cite{Rath2013,Levinsen2015,Guenther2018}. We moreover find that a smaller gas parameter  leads to deeper binding, reflecting that  the BEC becomes more compressible and hence induces a stronger effective interaction.

The predictions of our effective theory are in remarkably good agreement with the numerical DMC results for the entire considered range of coupling strengths $k_na$. This level of agreement is particularly striking in the strong-interaction regime, $k_na\gtrsim 1$, which does not offer a small parameter  to develop a controlled many-body theory. Yet, the predictive power of our description arises from the systematic combination of two reliable theories. First, 
the  boson-impurity scattering is treated within the ladder approximation, 
which has turned out to be surprisingly accurate for cold atomic gases~\cite{Strinati2018}. 
Second, the BEC density oscillations that mediate the  interaction are described by Bogoliubov theory, which is accurate for the typical situation of a small gas parameter. 
Respectively, our approach presents a rare instance of an intuitively simple yet accurate theory for a strongly interacting many-body system.

\begin{figure}
\begin{center}
\includegraphics[width=\columnwidth]{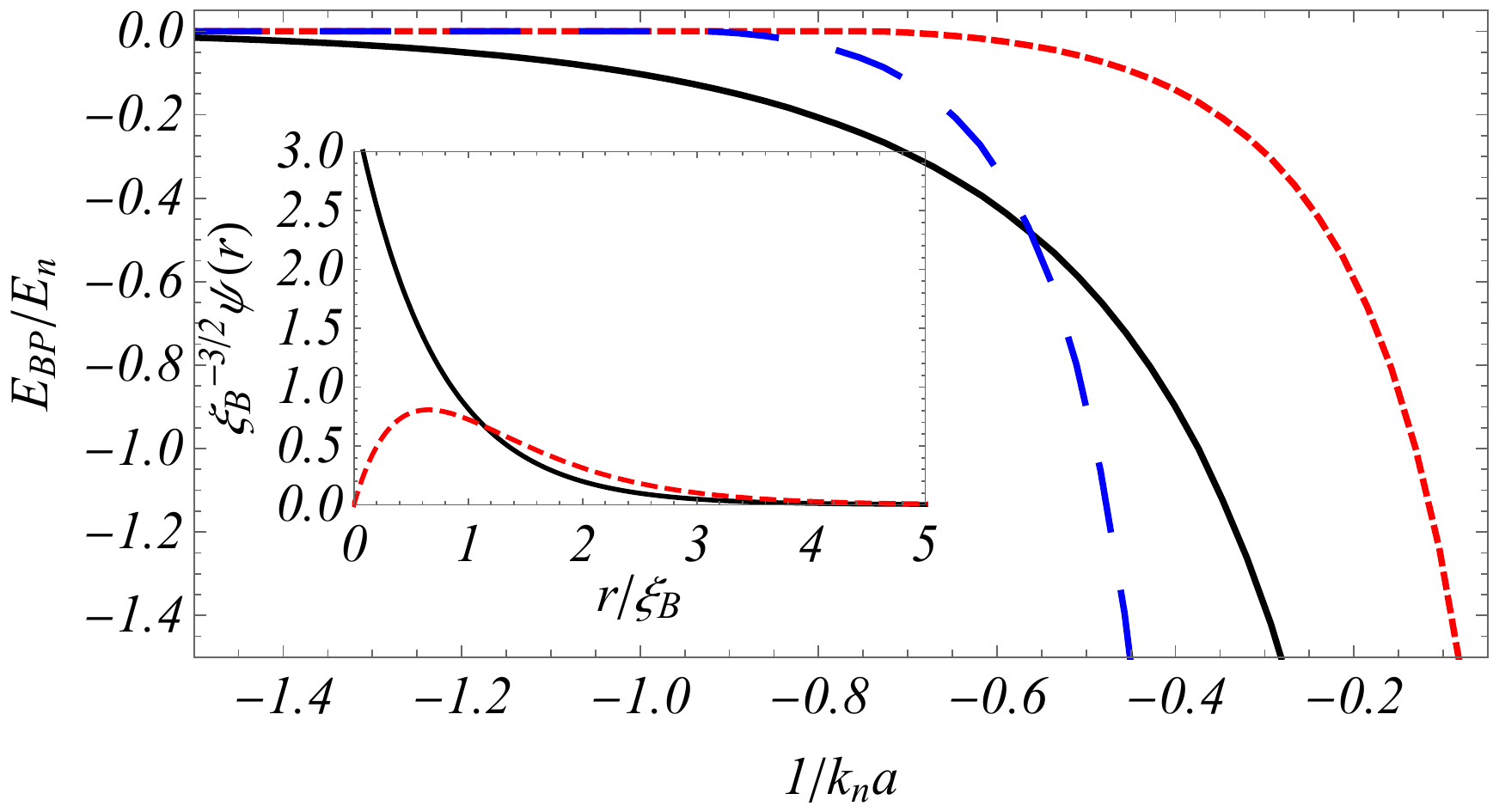}
\end{center}
\caption{Binding energy $E_{BP}$ of two bosonic (black solid line) and fermionic (red dashed line) impurities with the mass ratio $m/m_B=40/23$ for $n_Ba_B^{3}=10^{-6}$. 
The dashed blue line is to the Yukawa binding energy for the $p$-wave bipolaron.
Inset:  the radial parts of  the $s$- and $p$-wave functions (solid black and dashed red respectively) for $1/k_n a=-0.4$. }
\label{EnergyFermiBipolaron}
\end{figure}

In Fig.\ \ref{EnergyFermiBipolaron}, we compare the resulting bipolaron energy for the two cases of bosonic and fermionic impurities. We have chosen the mass ratio $m/m_B=40/23$ corresponding to the experimentally relevant case of $^{40}$K fermionic atoms in a $^{23}$Na BEC~\cite{Park2012,Zhu2017}.
While both cases promote the formation of bipolaron states beyond a critical interaction strength, Fig.~\ref{EnergyFermiBipolaron} clearly illustrates that fermionic 
impurities are more weakly bound that their bosonic counterparts. This is simply because their  wave function must have $p$-wave symmetry.  

To accurately determine  the critical coupling strength $k_na_c$ for bipolaron formation, we consider the size 
$\sigma=\sqrt{\langle r^2\rangle}/\xi_B$ of the  dimer state with $\langle r^2\rangle=\int\!d^3r|\psi({\mathbf r} )|^2r^2$. Since $\langle r^2\rangle$ diverges when the polarons unbind, the inverse $1/\sigma$ provides a clear indicator of the critical interaction strength. 
Indeed, its dependence on $1/k_na$ depicted in Fig.\ref{EnergyBoseBipolaron}(c) features a kink at $k_na_c$ beyond which $1/\sigma$ increases abruptly from zero.
Our theory recovers the classic results for the critical coupling strength   $\sqrt2/\alpha\xi_B m_r=1.1905$ and $\sqrt2/\alpha\xi_B m_r=0.2202$ for bound  $s$- and $p$-wave states in the Yukawa potential~\cite{Harris1962,Rogers1970,Edwards2017}. This demonstrates the accuracy of our approach.

\begin{figure}
\begin{center}
\includegraphics[width=\columnwidth]{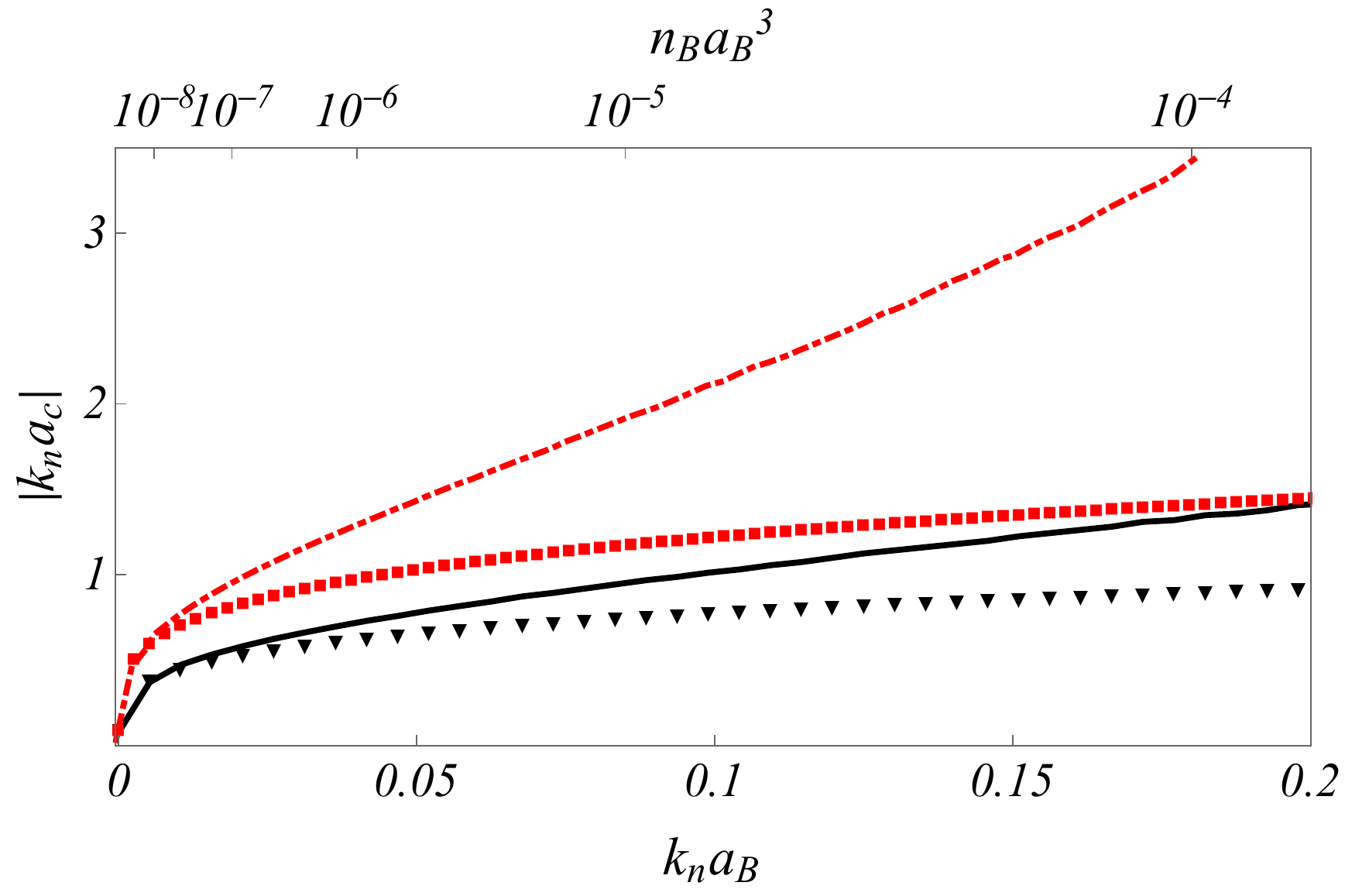}
\end{center}
\caption{The critical interaction strength $k_na_c$ for the formation of bipolarons as a function of $k_na_B$ (or $n_Ba_B^3$) for 
bosonic (black solid line) and fermionic impurities(red dashed line). Black triangles and red squares are the Yukawa result for bosonic and 
fermionic impurities respectively. 
}
\label{Correlation}
\end{figure}

The Yukawa interaction Eq.\ \eqref{Veffweak}, which results from a second order treatment within the Fr\"ohlich model, is accurate only for weak 
interactions $k_n|a|\ll1$. Indeed, it predicts  critical interaction strengths $k_n a_c$ and binding energies $E_\text{BP}$ substantially  
different from our strong coupling theory in Figs.\ \ref{EnergyBoseBipolaron} and \ref{EnergyFermiBipolaron}. 
This is because second order theory approximates  $\mathcal{T}({\mathbf k},0)\approx\mathcal{T}_\nu$, which is a significant overestimation for $k_n a\gtrsim1$.
Since the bipolaron is observable only for not too small interaction strengths, the Fr\"ohlich model is insufficient to analyse bipolarons in atomic gases. 
This is further illustrated in Fig.\ \ref{Correlation}, where we show the critical interaction strength $k_na_c$ as a function of the gas parameter $n_Ba_B^3$, 
obtained using both Eq.\ \eqref{Veff}, and the Yukawa potential Eq.\ \eqref{Veffweak}. As can clearly be seen, the Yukawa potential 
 is reliable only  for weak impurity-boson interaction, where the BEC has to be very compressible in order for the induced interaction 
  to bind  two polarons. 

The two Bose polaron experiments so far, which had the gas parameters $n_Ba^3\approx2\times10^{-8}$~\cite{Jorgensen2016} 
and $n_Ba^3\approx2\times10^{-5}$~\cite{Hu2016}, both used radio-frequency (RF) spectroscopy  to observe the polaron.
The same technique can in fact  be employed to detect bipolarons, 
whereby the RF field induces photo-association of polaron dimers leading to a resonantly enhanced atom-loss signal. In both measurements, the observed 
polaron spectrum had a typical line width of $\sim E_n$. The bipolarons found in our strong coupling theory should thus be observable for 
strong interactions, where we predict a bipolaron-resonance to emerge well separated from the single-polaron signal. 
 A natural question arises whether there are bound states  of more than two polarons, e.g. tripolarons consisting of three polarons. Indeed, it was found for the 
Yukawa potential that tripolarons can be stable, but only for a narrow range of coupling strengths and with a  small binding energy: At the threshold $k_na_c$ for 
bipolaron formation, the binding energy of the tripolaron is $-0.29 \,k_n a_BE_n$~\cite{Moszkowski2000} making them very hard to observe for $k_na_B\ll1$.
 We note that the attractive interaction mediated by Bogoliubov modes also can give rise to superfluid pairing in 
Bose-Fermi mixtures~\cite{Heiselberg2000,Viverit2002,Matera2003,Wang2006,Suzuki2008,Enss2009}.

In summary, we showed that two polarons formed by impurities in a BEC can merge into a bipolaron state that is bound by a nonlocal interaction mediated by phonons in the BEC. The  bipolaron states are a pure many-body effect arising from the surrounding BEC. They are therefore distinct from three-body Efimov states of two impurities and one boson, which are stable in a vacuum~\cite{Naidon2016}. 
 The theory described in this work opens the door for a number of future investigations. For example, the nonlocal nature of the effective interaction suggests exotic and interesting many-body physics of multiple interacting polarons. This question as well as the potentially profound effects of different system dimensions should be addressable in future work by the presented theoretical framework. We finally note that the induced interaction between Fermi polarons is rather weak \cite{Mora2010}, which has made the observation of bipolarons in degenerate Fermi gases  challenging~\cite{Scazza2017}. On the other hand, the results of this work show that the observation of bipolarons should now be possible in currently available BECs \cite{Jorgensen2016}, presenting an exciting positive outlook on future experiments.\\

We thank Jan Arlt and Pietro Massignan for valuable discussions. 
This work was supported by the Villum Foundation and the Danish National Research Foundation through a Niels Bohr Professorship.

\begin{widetext}
\section{Supplemental Material}
\subsection{Polaron quasiparticle properties}
The energy $\omega_{\mathbf k}$ and residue $Z_{\mathbf k}$ of a polaron with momentum $\mathbf{k}$ are given by 
\begin{align}
\omega_{\mathbf k}=\frac{\mathbf{k}^2}{2m}+\text{Re}\Sigma(\mathbf k,\omega_{\mathbf k}), \hspace{1cm}
Z_{\mathbf k}=\left(1-\frac{\partial\text{Re}\Sigma(\mathbf p,\omega)}{\partial \omega}\right)^{-1}_{\omega=\omega_{\mathbf k}},
\end{align}
 where $\Sigma(\mathbf p,\omega)$ is the impurity self-energy. We determine  $\Sigma(\mathbf p,\omega)$ using the diagrammatic scheme 
shown in Fig.\ \ref{StrongCouplingSelf}. 
\begin{figure}[!h]
\begin{center}
\includegraphics[width=5.1in]{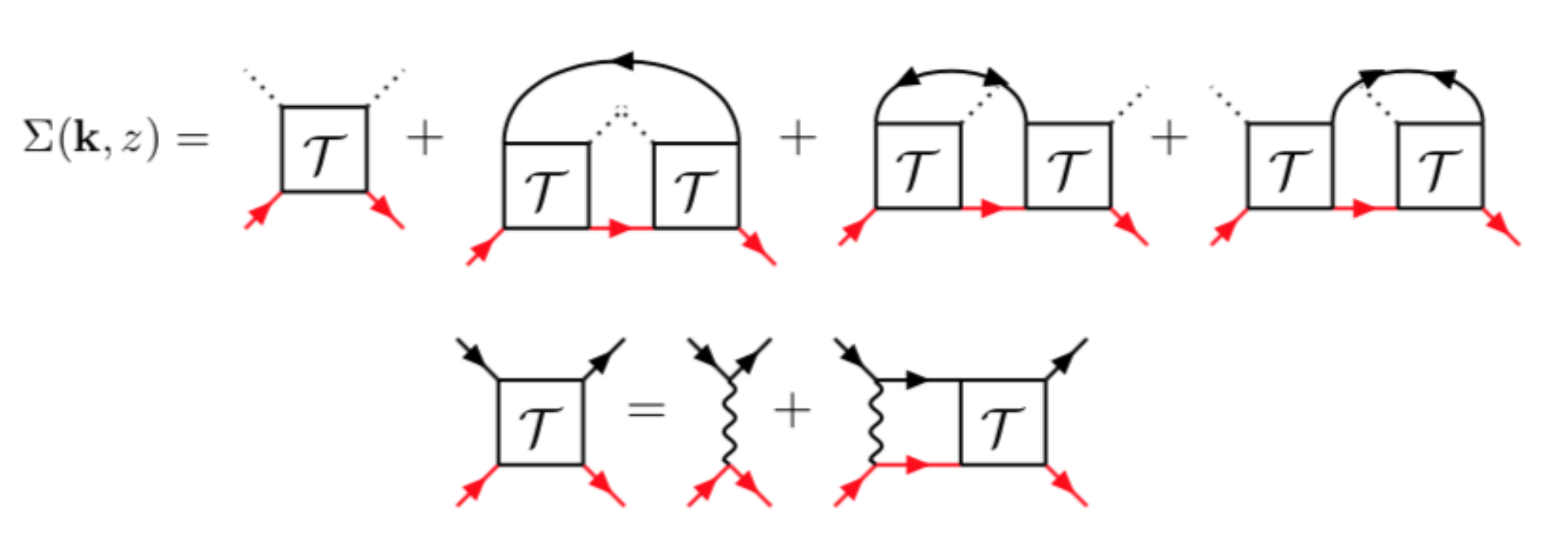}
\end{center}
   \caption{(Top) Self-Energy of an impurity coupled to the BEC. Red solid lines are the impurity propagator, solid black lines denote the BEC propagators,
    while the black dashed lines denote the condensate particles. (Below)We illustrate the ladder approximation for the boson-impurity scattering.}
   \label{StrongCouplingSelf}
 \end{figure}
This gives 
\begin{equation}
\Sigma(\mathbf k,z)=n_0\mathcal{T}(p)-n_0\sum_k G_{11}(k)\mathcal{T}^2(k+p)G(k+p)-2n_0\mathcal{T}(p)\sum_k G_{12}(k)\mathcal{T}(k+p)G(k+p),
\end{equation}
where $k=(\mathbf k,z)$ represents the energy-momenta vector, and $n_0$ is the condensate density. As we assume $T=0$ and a  weakly interacting BEC, we set  $n_0= n_B$.
The boson-impurity scattering matrix is calculated using the ladder approximation as 
\begin{align}
\mathcal{T}(\mathbf k,z)=\frac{\mathcal{T}_\nu}{1-\mathcal{T}_\nu\Pi_{11}(\mathbf k,z)},
\label{Tbi}
\end{align}
where $\Pi_{11}(\mathbf k,z)$ denotes the regularised pair propagator given by
\begin{align}
\Pi_{11}(\mathbf k,z)=\int\frac{d^3p}{(2\pi)^3}\left( \sum_{i\omega_{\nu}}G_{11}(\mathbf p,i\omega_{\nu})G(\mathbf k-\mathbf p,z-i\omega_\nu)+\frac{2m_\text{BI}}{p^2}\right).
\end{align}
 Here, $\omega_\nu=(2\nu+1)\pi T$ is a Fermi Matsubara frequency. 
 The BEC is described accordingly to  Bogoliubov theory, where the normal and anomalous BEC Green's functions are
 \begin{align} 
G_{11}(\mathbf{k},z)=\frac{u_{\mathbf{k}}^2}{z-E_{\mathbf{k}}}-\frac{v_{\mathbf{k}}^2}{z+E_{\mathbf{k}}}\hspace{1cm}
G_{12}(\mathbf{k},z)=\frac{u_{\mathbf{k}}v_{\mathbf{k}}}{z+E_{\mathbf{k}}}-\frac{u_{\mathbf{k}}v_{\mathbf{k}}}{z-E_{\mathbf{k}}}.
\end{align}
Here $E_{\mathbf{k}}=[\epsilon^B_{\mathbf{k}}(\epsilon^B_{\mathbf{k}}+2\mu_B)]^{1/2}$ is the 
Bogoliubov spectrum,   $\mu_B=4\pi a_Bn_B/m_B$ is the chemical potential of the bosons, and 
$u_{\mathbf{k}}^2/v_{\mathbf{k}}^2=[((\epsilon^B_{\mathbf {k}}+\mu_B)/E_\mathbf{k}\pm 1]/2$ are the usual coherence factors.  
 \subsection{Bethe-Salpeter equation and Schr\"odinger equation for two polarons}
From the scattering matrix $\Gamma(k_1,k_2;k_3,k_4)$ of two bare impurities, we obtain the scattering matrix for two polarons as 
$\Gamma_{\text{P}}(k_1,k_2;k_3,k_4)=Z_{\mathbf {k}_1}Z_{\mathbf {k}_2}\Gamma(k_1,k_2;k_3,k_4),
$ using the 
pole expansion for the impurity Green's function, $G(\mathbf k,z)=Z_{\mathbf k}/(z-\omega_{\mathbf k})$, $\Gamma_{\text{P}}$ also obeys a 
Bethe-Salpeter equation, but now with the effective polaron-polaron interaction 
 $V_{\text{eff}}(k_1,k_2;k_1-k_3)=Z_{\mathbf {k}_1}Z_{\mathbf {k}_2}V(k_1,k_2;k_1-k_3)$. Using the static approximation where the frequency dependence of the 
 interaction is neglected, the frequency sum in the Bethe-Salpeter equation only involves the 
 two impurity Green's functions and can be performed analytically. For zero center of mass, 
 the Bethe-Salpeter equation then reduces to the Lipmann-Schwinger equation, which in a compact matrix notation reads~\cite{Fetter1971}
 \begin{equation}\Gamma_P({\mathbf k},{\mathbf k}',E)=V_{\text{eff}}({\mathbf k},{\mathbf k}')+\sum_{\mathbf{ k}''}V_{\text{eff}}({\mathbf k},{\mathbf k}'')(E+i0_+-2\omega_{\mathbf{k}''})^{-1}\Gamma_P({\mathbf k}'',{\mathbf k}',E).
 \end{equation} 
 This equation describes the scattering of two polarons exchanging density oscillations in a BEC. Notice that $\Gamma_P$ represents the polaron-polaron
  scattering matrix and should not be confused with $\mathcal{T}$ given by \eqref{Tbi}, which gives the  impurity-boson scattering matrix. While the former includes
   the repeated polaron-polaron scattering with the mediated interaction  $V_{\text{eff}}({\mathbf k},{\mathbf k}')$, the latter refers to the
    ladder approximation of the impurity-boson  interaction.  
 
 The Lippmann-Schwinger equation is equivalent to a  Schr\"odinger equation for the relative wave function $|\psi\rangle$  of two polarons, 
 which   satisfies $\hat{V}_{\text{eff}}|\psi\rangle=\Gamma_P|\phi\rangle$. Here $|\phi\rangle$ is an eigenstate of the free Hamiltonian 
 such that $\langle \mathbf k|\hat{H}_0|\phi\rangle=\omega_{\mathbf{k}}\langle \mathbf k|\phi\rangle$, 
  $\langle \mathbf k |\hat{V}_{\text{eff}}|\mathbf k'\rangle=V_{\text{eff}}({\mathbf k},{\mathbf k}')$,
   and $\langle \mathbf k |\hat\Gamma_P|\mathbf k'\rangle=\Gamma_P({\mathbf k},{\mathbf k}')$. Therefore, the solution of the Lippmann-Schwinger equation for $|\psi\rangle$ coincides with the solution of the Schr\"odinger equation 
 
 \begin{equation}
E_{BP}\psi(\mathbf k)=2\omega_{\mathbf k}\psi(\mathbf k)+\sum_{\mathbf k'}V_{\text{eff}}(\mathbf{k},\mathbf{k}')\psi(\mathbf k').
\end{equation}

\subsection{Diffusion Monte-Carlo}
 The Hamiltonian of  two bosonic impurities with mass $m$ immersed in a gas of $N$ identical bosons with mass $m_B$ is given by
 \begin{eqnarray}
H&=&-\frac{1}{2m_B}\sum_{i=1}^N \nabla_i^2+\sum_{i<j}V_{BB}(r_{ij}) 
\label{Manyimpurities1}-\frac{1}{2m}\sum_{\nu=1}^2 \nabla_\nu^2
+\sum_{i=1}^N\sum_{\nu=1}^2V_{BI}(r_{i\nu}),
\nonumber
\end{eqnarray}
 where $r_{ij}=\left|\mathbf{r}_{i}-\mathbf{r}_{j}\right|$ and $r_{i\nu}=\left|\mathbf{r}_{i}-\mathbf{r}_{\nu}\right|$ denote the boson-boson and impurity-boson relative distance respectively. The boson-boson potential is modelled by a hard-sphere interaction where the diameter of the sphere is taken to be the scattering length $a_B$ of the BEC, 
\begin{equation}
V_{BB}(r)=\begin{cases}
+\infty & r< a_{B}\\
0 & r> a_{B} \;.
\end{cases}
\label{eq:Vbosons}
\end{equation}    
The impurity-boson scattering is modelled by an attractive square-well potential 
\begin{equation}
V_{BI}(r)=\begin{cases}
-V_0 & r< R_0\\
0 & r> R_0 \;,
\end{cases}
\label{eq:Vimp2}
\end{equation}  
characterised by a range $R_0$ and a depth $V_0$ given in terms of the impurity-boson scattering length\begin{equation}
a=R_0\left[ 1-\frac{\tan(K_0R_0)}{K_0R_0}\right] \;,
\label{eq:Vimp3}
 \end{equation}
where $K_0^2=2m_{BI}V_0$. For the attractive branch where there is no bound state between the impurity and a boson so that  $K_0R_0<\pi/2$.
 Here, we consider values of the range $R_0$ small compared to the boson-boson scattering length $a_B$ which again is small 
 compared to the interparticle distance, $R_{0}<a_{B}\ll n_{B}^{-1/3}$. These potentials were used for the  study of single polarons in Ref.\cite{Ardila2015}.

In the diffusion Monte-Carlo simulations we use $N+2$, $N+1$  and $N$ particles to determine the  energy of the bipolaron, polaron and BEC system respectively. 
Our calculations are performed in a cubic box of size $L$ with periodic boundary conditions. 
 The trail wave function $\psi_{T}(\mathbf R)=\Psi_B(\mathbf{R}_B)\Psi_I(\mathbf{R}_I) \Psi_{BI}(\mathbf{R}_B,\mathbf{R_I} )$ is written in terms of Jastrow functions
\begin{eqnarray}
\Psi_B(\mathbf{R}_B)=\prod_{i<j}f_{B}(r_{ij}), \hspace{0.5cm}
\Psi_I(\mathbf{R}_I)=\prod_{\alpha<\beta}f_{I}(r_{\alpha\beta}), \hspace{0.5cm}
\Psi_{BI}(\mathbf{R}_B,\mathbf{R}_I)=\prod_{i}^{N}\prod_\alpha^{2} f_{BI}(r_{i\alpha}),
\end{eqnarray} 
which are determined by the two-body solutions of the hard sphere and square well potentials.  
Finite size effects are analysed by changing the number of particles from 32 to 128 while increasing the size of the box to ensure that the density of the BEC remains fixed. The ground state energies are  obtained by propagating the Schr\"odinger equation in imaginary time $\tau= i t $. For the attractive branch of the polaron and bipolaron, this method provides the exact ground state energy.For technical details, see Ref.\cite{Ardila2015}.

\end{widetext}
\bibliography{Bipolaron} 

\bibliographystyle{apsrev4-1}

\end{document}